# *Real-time observation of the coherent transition to a metastable emergent state in $1T\text{-}TaS_2$.*

Jan Ravnik[1], Igor Vaskivskyi[1,2], Tomaz Mertelj[1,2], and Dragan Mihailovic[1,2]

[1]Jozef Stefan Institute, Jamova 39, SI-1000 Ljubljana, Slovenia

[2]CENN Nanocenter, Jamova 39, SI-1000 Ljubljana, Slovenia

The transition to a hidden metastable state in $1T\text{-}TaS_2$ is investigated in real time using coherent time-resolved femtosecond spectroscopy. Relying on spectral differences between phonon modes in the equilibrium states and in the metastable state, and temperature-tuning the metastable state lifetime, we perform stroboscopic measurements of the electronic response and switching of coherent oscillation frequency through the transition. Very fast coherent switching of the collective mode frequency is observed ($\sim$ 400 fs), comparable to the electronic timescale $\sim$300 fs. A slower, 4.7 ps process is attributed to lattice relaxation. The observations are described well by a fast electronic band structure transformation into the metastable state, consistent with a topological transition.

Metastable many-body states are manifestations of collective effects which are of great importance in different areas of science, from the false vacuum state in field theory of elementary particles to potential uses in memory devices. The microscopic mechanisms leading to such metastable states, and particularly the transition dynamics are still relatively unexplored. In condensed matter systems exhibiting electronic order, a transient particle-hole imbalance may lead to the creation of transient emergent states, whose structure may be different than any equilibrium states, as illustrated by a recent example of photoexcited "hidden" (H) state in the layered dichalcogenide 1T-TaS$_2$[1-3] (Fig. 1a). This material is an extremely versatile model system which has experienced a resurgence of recent interest because it exhibits a number of equilibrium and non-equilibrium phases[1,3] (Fig. 1b) where conventional ordering mechanisms do not apply. Competing Fermi surface instabilities, strain and Coulomb repulsion in combination with out-of-plane "orbitronics"[4] lead to a multidimensional phase diagram which includes a Mott insulator phase, a metallic phase, and a superconducting phase. A quantum spin liquid-like state was reported below 220 K, adding to its complexity[5]. One of the consequences of competing interactions is that it has a sequence of charge density wave (CDW) transitions starting with a transition from a metallic state above 550K to an incommensurate (IC) CDW state, followed at $T_{c2} = 350\ K$ by an unusual strain-driven transition[6] exhibiting a partial Kohn anomaly [7-9] to a "nearly commensurate" (NC) state composed of domain walls (discommensurations) and commensurate domains. This unusual transition, whose topological nature was discussed already by McMillan[7] is of great recent interest, as it is believed to play a role in metastable state formation[1,10]. Eventually the discommensurations disappear below $T_{c3} \simeq 180\ K$ in the transition to an insulating commensurate (C) charge-ordered state. On heating from the C state, transition to the trigonal (T) phase takes place at $T_{C-T} = 220\ K$, which survives up to 280 K. We are concerned here with the mechanism for the transition between the insulating C state and a metallic metastable hidden (H) state which takes place under non-equilibrium conditions. Initial experiments showed that short optical pulses may trigger such a transformation, but there is so far little indication on what timescale the new state settles into long range order. The fundamental issue is whether the transition takes place diffusively or coherently. In the diffusive case, carriers from the C ground state are delocalized by the short, 30 fs pulse whereupon they first localize in pseudo-random positions, and later form the textured domain structure[2] through a combination of diffusive electron motion[11], incoherent atomic ordering[1], followed by domain wall nucleation and growth on a 300-400 ps timescale as observed by X-ray diffraction at the NC-IC transition[12]. Alternatively, in a coherent transition, the transition to a long-range ordered metastable state takes place on an electronic timescale, starting with a many-body interference of nested non-equilibrium Fermi surface electrons followed by an adjustment to the lattice. In this case, the lattice is expected to adapt rapidly to the new electronic order forming a new periodic lattice distortion without any intermediate diffusive processes. Secondary relaxation of strain between the incommensurate electronic order and the underlying crystal lattice would then follow on much longer timescales. Another important question is what is the difference between the H state and a supercooled NC state[13], which is also related to the fundamental mechanism of formation of the H state.

The problem with investigating long-lived metastable states created by photoexcitation is that the transition is caused by a singular event – such as a single shot laser pulse. The state of the system before and after the transition can be investigated in detail by equilibrium techniques[1,2], but not the dynamics of the event itself. The H metastable state lifetime in 1T-TaS$_2$ has recently been discovered to be tunable by a combination of substrate strain and temperature[14,15], so provided the system can be made to relax in between laser pulses, stroboscopic pump-probe measurements of the collective and single particle dynamics *through* the C-H transition can be performed. While at 4K the extrapolated H state lifetime is $> 10^{18}$ s, at 160 K the lifetime is $\tau_H < 10^{-4}$ s, and the relaxation of the photoinduced state is sufficiently fast to allow measurements with a 1kHz repetition rate laser. At this base temperature, photoexcitation with pulse fluences of $\sim$ 1 mJ/cm² leads to heating of both the electrons and the lattice well above $T_{C-T} = 220$ K, so we might expect that either the H state, the NC or T state, or possibly some other nonequilibrium state might form[3]. Distinguishing the ordering vectors[2] between the H and NC states requires currently unavailable resolution with time-resolved electron diffraction or X-ray techniques[12,16], so the ordering dynamics cannot be measured directly by diffraction techniques. One possibility to discern between coherent and incoherent domain wall dynamics would be to investigate the linewidth of the mode commonly assigned to the amplitude mode (AM) (whose displacements are shown in Fig. 1c), similarly as was previously used to measure the coherent evolution of the order parameter through the CDW transition in TbTe$_3$ [17]. Alternatively, here we show that one can measure the apparent shift of the AM frequency associated with the transition[1] in real time using coherent phonon spectroscopy with multiple laser pulses.

To do this, we need to determine if the H state can be unambiguously distinguished from other states by its phonon spectrum, and particularly from the supercooled NC state which has been reported in thin flakes[10]. In Fig 2a we show the coherent reflectivity oscillations obtained with pump-probe spectroscopy using 30 fs, 800nm laser pulses at different temperatures. Fourier transformed phonon spectra together with fits to three sets of coherent oscillations by the displacive excitation mechanism using Eq. 1 from [18]) are shown in Fig. 2b. We see that the spectra in the four phases differ substantially. A summary of the T-dependences of our observed phonon peak positions in different (equilibrium) phases is shown in Fig. 2c, including Raman data (the Billouin-zone center phonons $q \simeq 0$)[19]. To show the origin of the soft modes, we also include neutron [20] and X-ray frequencies at the wavevector of the observed Kohn anomaly ($q = 0.26$ r.l.u.) [9] at the IC-NC transition. In the C state, the low frequency spectrum below 4 THz (120 cm$^{-1}$) consists of Ta modes derived from folding of high-temperature acoustic modes into the commensurate Brillouin zone[19]. Above 220K, in the IC and NC states, modes appear in the Raman spectra whose origin can be traced to the Kohn anomaly near the NC-IC transition seen in the X-ray and neutron data, and a number of soft modes associated with the transition can be identified at low temperatures[19]. A softened longitudinal acoustic (LA) mode associated with the observed Kohn anomaly at $q = 0.28$ r.l.u. along the crystal axes modulates the parent 1T structure to give the IC phase[9]. An accompanying softened transverse acoustic $TA_{||}$ mode induces the rotation of the overall ordering vector to accommodate the ordering of commensurate domains and discommensurations in the NC state. The NC structure, and similarly the H structure[2] can be described in terms of a superposition of different wavevectors (harmonics). Phonons whose wavevectors correspond to these harmonics may partially condense or soften in different points of the Brillouin zone, giving multiple collective modes in the folded Brillouin zone[19,21]. In the C phase, as the discommensurations disappear, and the collective "amplitude" mode becomes dominant in the spectrum (Fig. 1c). The intensity of the coherent oscillation (and Raman) spectra reflect the mode coupling to the electronic modulation, and the photon resonances for displacive excitation and detection as described by the stimulated Raman tensor[22]. The resonances depend on the electronic structure in the vicinity of the Fermi level, and are particularly sensitive to orbital ordering and change of symmetry[23]. This can be easily understood if we write the Raman tensor in terms of derivative of the complex dielectric constant $\boldsymbol{\varepsilon}(\omega)$ with respect to the phonon displacement $Q$ : $\boldsymbol{R}(\omega) = d\boldsymbol{\varepsilon}(\omega)/dQ$, where $\boldsymbol{\varepsilon}(\omega)$ depends on the joint density of electronic states involving optical transitions between states near $E_F$ and 1.5 eV above or below it[24].

Here we shall focus on the spectral differences obtained by pump-probe spectroscopy in the vicinity the AM[19] at 2.3 THz in the C, H and NC states (Fig. 2d). For the H state measurements, an initial switching pulse preceding the pump-probe pulses a few hundred ps was used to transiently drive the system into the H state. Comparing the spectra near the AM frequency in equilibrium phases with the H phase, we see quite remarkable differences (Figs. 2b and d). Three modes (at $\omega_1$, $\omega_2$ and $\omega_3$) are observed in all four phases (Fig. 2b). However, their intensities are drastically different, and each phase has a unique spectral fingerprint. The differences between the H and NC spectra are particularly remarkable. The origin of the spectral differences may be attributed to the high sensitivity of the low-energy band structure to orbital order[4], so differences in the long-range domain order may also result in different resonant coherent phonon intensities.

In addition to in-plane domain ordering, calculations by Ritschel[4] showed that different stacking of the charge-density wave along the c axis can lead to either metallic or insulating phases, depending on the Ta $5d$ "orbital textures". Band structure calculations show that the stacking appears to have a very strong effect not only on out-of-plane electronic bands, but also on in-plane bands near $E_F$. Since the AM displacements predominantly involve Ta atoms (Fig. 1c), the electron-phonon coupling is strongly dependent on the Ta 5d orbital texture. This can lead to further phonon mode differences in frequency and intensity between the H, C and NC phases.

Having established that the AM frequency can be used as a fingerprint of the H state, we perform a stroboscopic three-pulse experiment at 160 K, where the metastable H-state lifetime (< 100 $\mu$s) is much shorter than the time between pulses (1 ms)[14]. To determine the real-time trajectory through the C-H transition on short timescales first, a chopped beam of Pump (P) pulses is applied starting a coherent oscillation of the collective modes, which are recorded by reflectivity oscillations $\Delta R(t)$, recorded by lock-in amplifier and a variably time-delayed probe (p) pulse. (A schematic diagram of the experiment is shown in Fig. 3a.) The H state transition is induced at variable delays after the P pulse by a "destruction" pulse (D). Fluence of the D pulse is adjusted above and below the switching threshold (~1 mJ/cm²) while the P and p fluences are kept constant at 0.2 and 0.05 mJ/cm² respectively. The direct change of the reflectivity, $\Delta R_\mathrm{D}$, due to the D pulse induced transition and the change in the frequency of the P-pulse-induced coherent oscillations induced by the transition are recorded. The results are shown in Fig. 3 b and c above and below the switching threshold[1] as a function of time-delay $\tau_\mathrm{P-D}$ between the P and D pulses. The fluences of D pulse were set above and below the switching treshold at 1.5 and 0.7 mJ/cm², respectively. The transition caused by the D pulse is clearly observed from the change of the coherent oscillation pattern. To analyze the change quantitatively we make an expanded plot at one $\tau_\mathrm{P-D}$ delay (shown superimposed on the 2D plot in Fig. 3b) and focus only on the oscillatory signal by subtracting exponentially decaying components of the transient reflectivity (Fig. 3e). We clearly see a frequency change in the magnified plot of the coherent oscillations which takes place in a time interval between $\tau_{\text{P}-\text{p}} = 2.4$ ps and 2.8 ps on the time axis. Fitting the oscillations with a function of the form $\sim\sin[\omega(t)\,t]$, reveals a step-like change of the collective mode frequency from 2.31 to 2.23 THz within $\tau_t \sim 400$ fs (top trace, Fig. 3e), followed by a small frequency shift on a timescale of $4.7 \pm 1$ ps reaching the long-time value of $\omega_H = 2.25$ THz. The frequency transition time $\tau_t$ corresponds approximately to one mode oscillation period ($T_{\text{AM}}^{\text{C}} = 0.44$ ps). Below the switching threshold (Fig. 3c) ($F_{\text{D}} = 0.7$ mJ/cm$^2$), no transition is observed, and as expected, there is no measurable change of the frequency since the initial and final states are the same.

The C-H transition is accompanied by an increase of reflectivity of $\Delta R_{\text{D}} \sim 5\,\%$ at 800 nm with a risetime $\tau_r \lesssim 100$ fs, consistent with observed melting of the C state[25]. Thereafter, the reflectivity relaxes with two characteristic lifetimes ($\tau_1 \simeq 300 \pm 20$ fs and $\tau_2 \simeq 4.7 \pm 0.5$ ps). The two transients of the frequency and reflectivity switching, given by $\tau_t = 400$ fs and $\tau_e = 300$ fs, respectively, characterize the initial transition followed by processes on a timescale of $\sim 4.7$ ps, previously attributed to the lattice relaxation at the IC-NC transition[26].

To understand the observed behavior of the coherent AM oscillation through the C-H transition, and the meaning of the observed lifetimes, we need to consider the sequence of ordering events leading to the eventual creation of the H state. In equilibrium, at high temperatures, the single band metallic system[27] is susceptible to a transition to an incommensurate state driven by Fermi surface (FS) nesting which occurs at 550K [28]. In the photoexcited state however, experimentally 1$T$-TaS$_2$ shows a significant $\text{e} - \text{h}$ asymmetry [29], indicating a shift of the chemical potential (photodoping) and a warped photoexcited FS. On the timescale of $300{\sim}400$ fs that the transition takes place, the estimated effective electronic temperature given by the two-temperature model is possibly $> 1000$ K [1], [29] (the estimated temperatures are shown in Fig. 3f). While we should not take this temperature to be very accurate, the implication is that the ordering of nested FS electrons takes place outside of equilibrium. At the end of this process, we can conclude that the resulting transient CDW state will have a modulation wavevector which is different from the equilibrium one.

The next step is the mutual adjustment of the non-equilibrium electronic order and the lattice. The observed $\tau_2 \sim 4.7$ ps (Fig. 3d) is consistent with the time for the formation of a periodic lattice distortion in the NC state measured by ultrafast electron diffraction[26,30]. It is also consistent with the thermalization time $\tau_{th}$ of the electrons with the lattice given by the two-temperature model ($\tau_{th} = 4 \sim 5\ ps$) shown in Fig. 3f. $\tau_2$ is also consistent with the anharmonic lifetime $\tau_{anh} \sim 5.9$ ps given by the Lorentzian linewidth of the AM in the H state (0.17 THz) which signifies the time needed for energy of the AM to be released to other coupled lattice modes. After this time, the accommodation of remaining strain due the incommensurability between the transient electronic order and the lattice leads to the formation of commensurate domains and discommensurations in the new metastable H state.

To model the transition quantitatively, we first attempt to use a model based on a time-dependent Ginzburg-Landau theory for a first order transition with a time-dependent potential to describe the metastable state shown schematically in Fig. 4a. The timescale for the formation of the metastable state indicated by experiments are the electronic $\tau_1$, followed by lattice ordering on a timescale of $\tau_2$. The metastable state naturally arises through a third-order term in the free energy expansion $F = \alpha(t)\psi^2 - \beta(t)\psi^3 + \gamma\psi^4$, as allowed by symmetry of the triple CDW in 1$T$-TaS$_2$[31], where $\psi$ corresponds to the dimensonless order parameter. The time-dependence in the potential[32] can then be introduced through experimental decay constants in the form: $\alpha(t) = \alpha_0[1 - \exp(-t/\tau_1)]$, $\beta(t) = \beta_0[1 - \exp(-t/\tau_2)]$. Numerical simulations of the transient reflectivity based on equations of motion derived from this model, under the assumption that the transient reflectivity is proportional to the order parameter $\Delta R \sim \Delta(\psi^2)$ [32] is shown in Fig. 4b. Even with significant adjustment of the parameters, agreement with the experimental data in Fig. 3b is hardly satisfactory, showing that modelling in terms of an order parameter trajectory through an ordering transition that was used successfully to describe TbTe$_3$ [32] does not work here.

Following the notion that the transition cannot be described using Landau theory with a conventional order parameter, we consider the effect of the change of electronic structure of states near $E_F$ associated with the insulator-metal topological transition on the collective mode spectrum. To describe the frequency shift phenomenologically (which includes the effect of a resonant crossover between $\omega_2$ and $\omega_3$), we can write $\omega = \omega_C - \delta(t)$. We can then calculate the transient reflectivity dynamics of the AM through the transition using a simple quadratic potential $F = \frac{\omega^2(t)}{2}\psi^2$ with an equation of motion $\frac{1}{\omega^2}\psi'' + \frac{\alpha}{\omega}\psi' + \psi = 0$, where to first order $\omega^2(t) = \omega_C^2 - 2\omega_C\delta(t)$, and $\delta(t) = \frac{1}{2}\left[\eta_1 + \eta_2 e^{-t/\tau_1} + \eta_3 e^{-t/\tau_2}\right]\left[1 + \text{erf}\left(\frac{t}{\tau_r}\right)\right]$. The predicted transient reflectivity response within this model is very sensitive to the fast time-evolution of the overall frequency trajectory. In Fig. 4d we show the predicted behavior with an electronic timescale transition to the metastable state: $\tau_1 = 300$ fs only ($\eta_3 = 0$) with a finite long-delay frequency shift $\eta_1$. For comparison, the reflectivity for $\eta_1 = 0$ is shown in Fig. 4e. We see that in both cases the model compares well with the observed transient reflectivity data in Figs. 3b and c, above ($\eta_1 > 0$) and below ($\eta_1 = 0$) the switching treshold. Adding the third term ($\eta_3 \neq 0$) with $\tau_2 = 4.7$ ps in $\delta(t)$ gives a barely perceptible change in the predicted response, from which we conclude that the change of $\omega$ is predominantly the result of electronic ordering into the H state, while the lattice response which occurs on a timescale of $\tau_2$ has little effect on the AM.

The use of coherent mode dynamics through the transition highlights the electronic nature of the transition driving the ordering in the H state on a timescale $< 400$ fs. We conclude that the C-H transition cannot be described in terms of a time-evolution of a Landau-order parameter through a first order transition. Rather, it is consistent with a non-equilibrium topological transition from a uniform state to a textured state with an accompanying change of electronic structure giving rise to good agreement with the phonon mode dynamics through the transition. The marked difference in the H and NC state mode spectra is quite surprising, implying that their electronic structure near the Fermi level is substantially different, in spite of the fact that the both are composed of long-range order of commensurate domains and discommensurations which look qualitatively similar in scanning tunneling microscopy[2]. Understanding the origin of the difference in electronic structure of different states and in particular between H and NC states will be important for establishing the microscopic mechanism promoting metastability of the transient emergent state.

**Acknowledgments**

We acknowledge discussions and comments from S. Brazovskii. Authors would like to acknowledge the financial support from ARRS, P1-0040, ERC ADG GA 320602 Trajectory and ERC PoC CDWmem.

**Figure captions**

Fig. 1 (a) The structure of 1T-TaS$_2$. (b) Schematic phase diagram of 1T-TaS$_2$ upon cooling, warming and after photoexcitation. (c) A schematic diagram of the atomic displacements corresponding to the dominant breathing "amplitude mode" in the C state.

Fig.2 (a) Time-resolved reflectivity of 1T-TaS$_2$ in different states, which show exponential quasiparticle decay with superimposed oscillatory response from coherent excitation of collective modes and (b) their fast Fourier transforms. Dots are the experimental data; thick lines are fits with the DECP model[18]. Thin lines are individual oscillators used for the DECP fit. The transition from the C to the H state results in a shift of intensity from $\omega_1$ to $\omega_2$. In the NC state $\omega_3$ is dominant. (c) Temperature dependence of the collective mode frequencies combined from different techniques[9,19,20] shown in the legend. (d) Temperature dependence of the mode frequencies. The size of the symbols and color intensity of the streaks represent the intensity of the corresponding modes in the C, H, NC and T phases in blue, red, cyan and light cyan respectively.

Fig. 3 (a) Schematic diagram of the optical three-pulse experiment. (b) and (c) Three-pulse P-D-Pr traces for D pulse fluence adjusted above (F$_D$=1.5 mJ/cm$^2$) and below (F$_D$=0.7 mJ/cm$^2$) switching threshold respectively. Dashed line shows the time of D pulse arrival. Black curve is a single trace for selected τ$_{P-D}$. (d) Standard two-pulse pump-probe trace with above-threshold pump pulses (F$_P$=1.5 mJ/cm$^2$). The reflectivity of the sample possesses fast rise and double-exponent relaxation. (e) Detailed analysis of the P-D-Pr trace for τ$_{P-D}$=2.6 ps (black) and extracted oscillatory response only (red dots) with FM sine fit (red line). The frequency profile used for the fit is shown in purple. Bottom panel: the same data zoomed-in around τ$_{P-p}$=τ$_{P-D}$. Dashed lines are sine fits before (blue) and after (green) the D pulse arrival using constant frequency $\omega_C$ and $\omega_H$ respectively. The narrow region, in which the phase shift occurs and both fits diverge from data is highlighted. (f) Estimation of electron and lattice temperature after D pulse arrival using two temperature model for $F_D$=1.5 mJ/cm$^2$ [1].

Fig. 4 (a) Time evolution of anharmonic potential used for simulations with Ginzburg-Landau theory. Starting from global minimum on free energy surface (1) the D pulse drives the system to the high-symmetry state (2), from which it thermalizes via intermediate states (3) to the new metastable state, which correspond to the local minimum on the equilibrium free energy surface (4). (b) Simulated three-pulse spectrum using a Ginzburg-Landau model (a). (c) Evolution of quadratic potential used for the topological model. Starting from the ground state (1), the system follows to the transient state with high concentration of discommensurations (2), which then decrease to the new, non-zero value (3). [1] (d) and (e) Results of simulations with model from (c) for $F_D$ above and below the switching threshold respectively.

Fig. 1

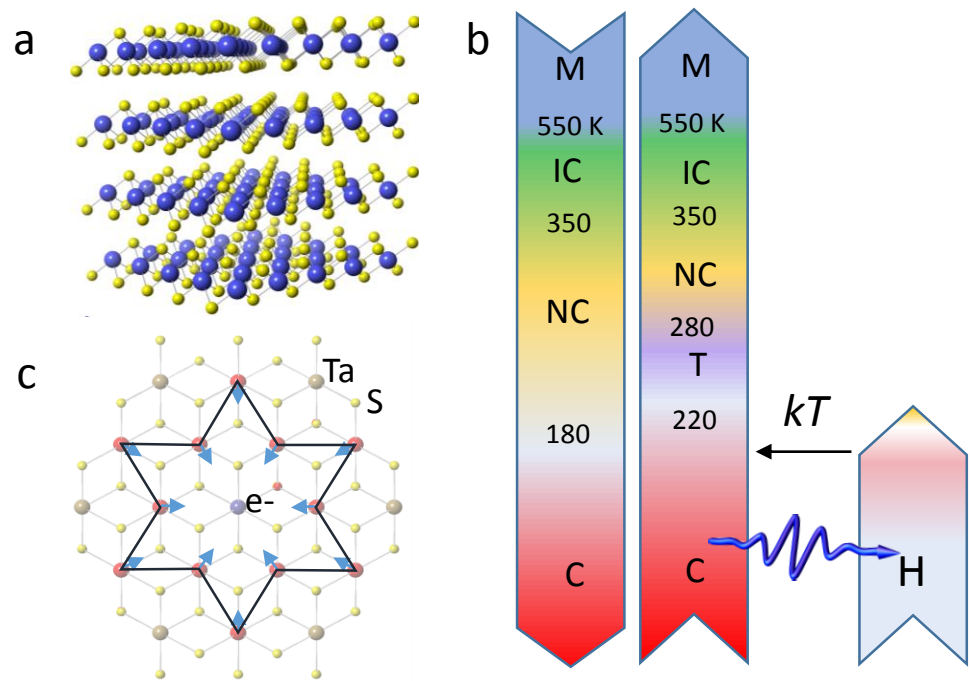

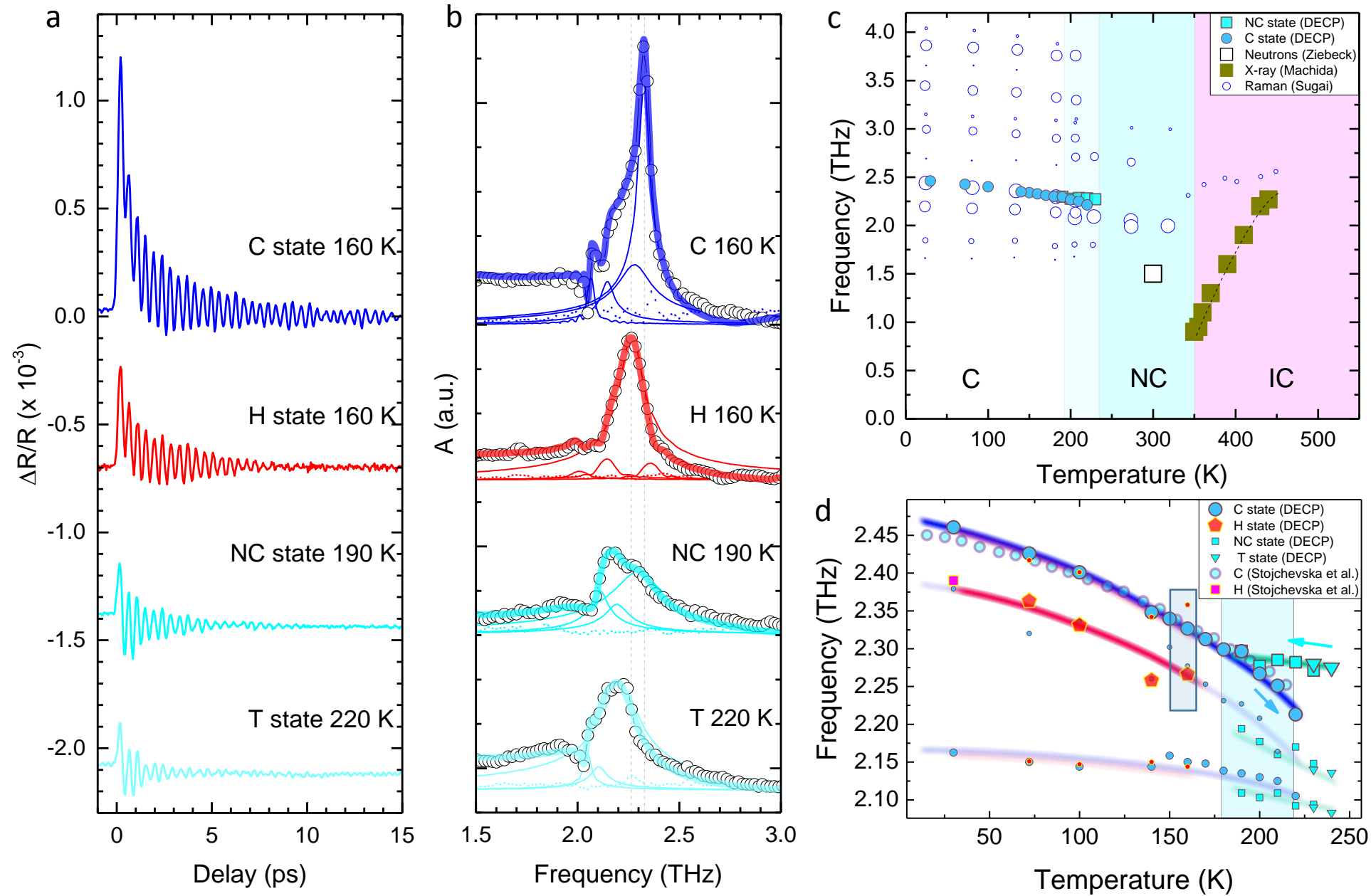
Fig. 2

Fig. 3

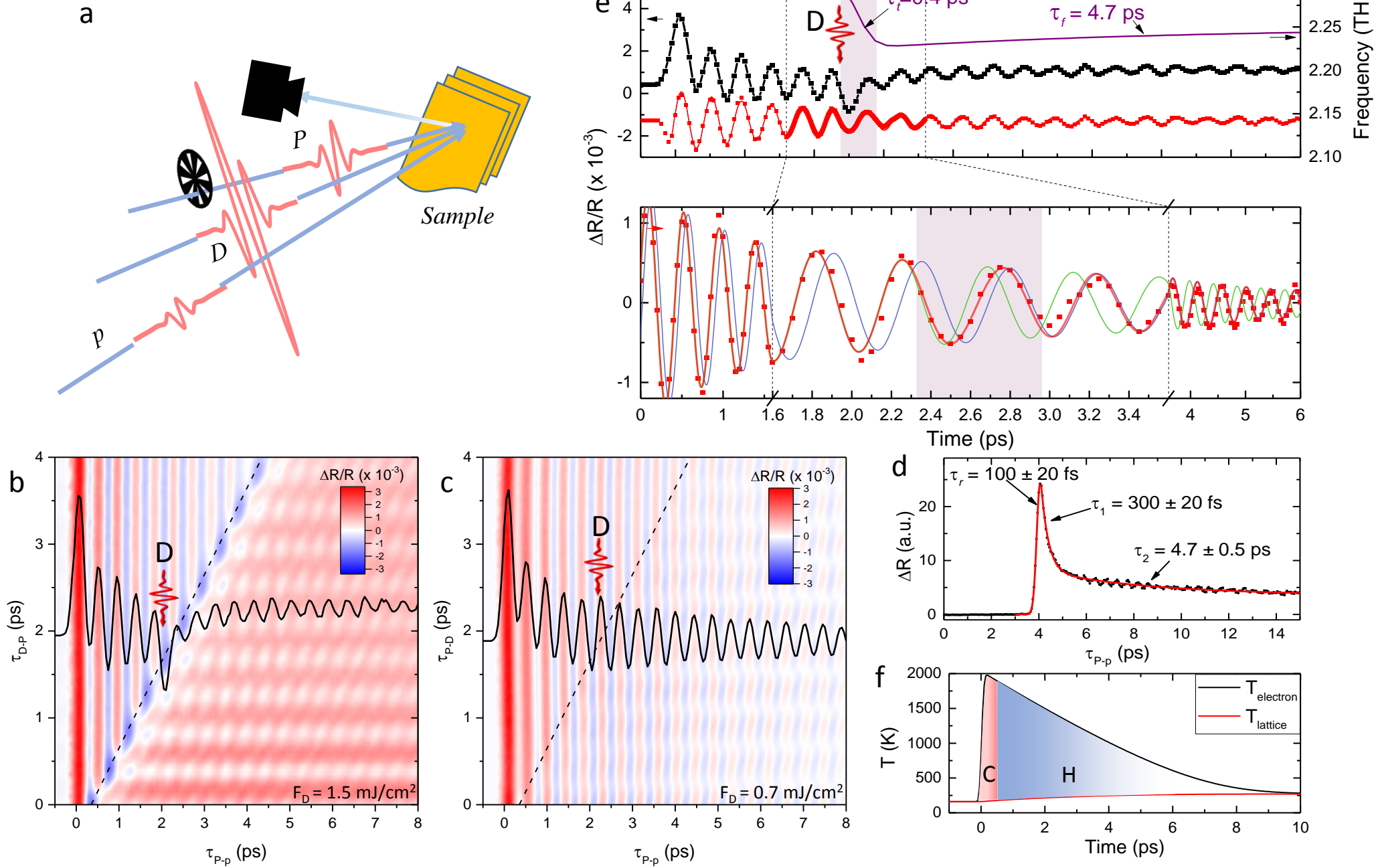

Fig. 4

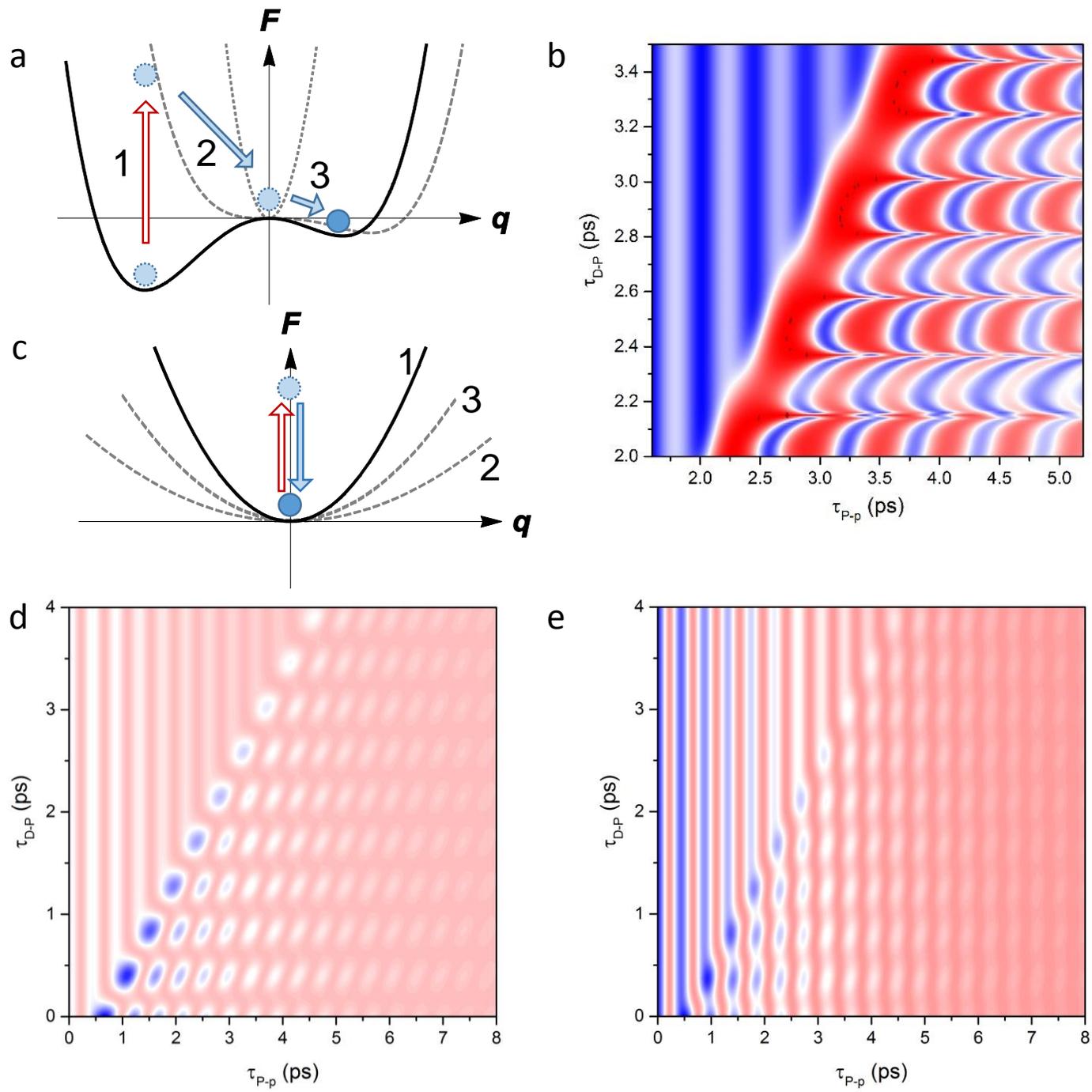